\documentstyle[preprint,prl,aps]{revtex}
\begin{document}
\def\sn2{$\sin^22\theta$}
\def\dm2{$\Delta m^2$}
\def\ch2{$\chi^2$}
\def\ltap{\ \raisebox{-.4ex}{\rlap{$\sim$}} \raisebox{.4ex}{$<$}\ }
\def\gtap{\ \raisebox{-.4ex}{\rlap{$\sim$}} \raisebox{.4ex}{$>$}\ }

\newcommand{\EQ}{\begin{equation}}
\newcommand{\EN}{\end{equation}}
\draft

%%%%%%%%%%%%%%%%%%%%%%%%%%%
% General purpose symbols %
%%%%%%%%%%%%%%%%%%%%%%%%%%%

\def\bec{\begin{center}}
\def\eec{\end{center}}

\def\beq{\begin{equation}}
\def\eeq{\end{equation}}

\def\degres{\mbox{$^\circ$}}      % mathematical symbols

\def\gsim{\mbox{$\stackrel{_>}{_\sim}$}} %Maggiore o circa%
\def\lsim{\mbox{$\stackrel{_<}{_\sim}$}} %Minore   o circa%

%
%Redefinition of fraction simbols of general use
%
\newcommand{\stdfrac}[2]{{{#1}\over{#2}}}
\renewcommand{\frac}[2]{{{\displaystyle #1}\over{\displaystyle #2}}}

\def\eV{\mbox{ eV}}      % general quantities
\def\keV{\mbox{ keV}}
\def\MeV{\mbox{ MeV}}

\def\BeSv{\mbox{$^{7}$Be}}     % Isotopes
\def\BHt{\mbox{$^{8}$B}}

\def\calA{\mbox{$\cal A$}}     % Calligraphic letters
\def\calB{\mbox{$\cal B$}}
\def\calC{\mbox{$\cal C$}}
\def\calH{\mbox{$\cal H$}}
\def\calM{\mbox{$\cal M$}}
\def\calN{\mbox{$\cal N$}}
\def\calO{\mbox{$\cal O$}}
\def\calR{\mbox{$\cal R$}}
\def\calS{\mbox{$\cal S$}}

\def\hatg{\mbox{${\hat{g}}$}}   % hat symbols
\def\hath{\mbox{${\hat{h}}$}}
\def\hats{\mbox{${\hat{s}}$}}
\def\hatt{\mbox{${\hat{t}}$}}

\def\dms{\mbox{$\Delta m^2$}}     % neutrino MSW parameters
\def\dMs{\mbox{$\Delta M^2$}}

\def\SdTv{\mbox{$\sin2\theta_V$}}
\def\SdTvS{\mbox{$\sin^2 2\theta_V$}}
\def\STvS{\mbox{$\sin^2 \theta_V$}}
\def\STv{\mbox{$\sin \theta_V$}}

\def\DeltaMs{\mbox{$\Delta m^2$}}
\def\ThetaV{\mbox{$\theta_v$}}
\def\ThetaM{\mbox{$\theta_m$}}

\def\GF{\mbox{$G_F$}}               % Fermi constant

                                    %
                                    % neutrino symbols
\def\numt{\mbox{$\nu_{\mu(\tau)}$}} %  nu_mu(tau) neutrinos
\def\nue{\mbox{$\nu_e$}}            %  nu_e
\def\num{\mbox{$\nu_\mu$}}          %  nu_mu
\def\nut{\mbox{$\nu_\tau$}}         %  nu_tau
\def\nuone{\mbox{$\nu_1$}}          %  nu_1
\def\nutwo{\mbox{$\nu_2$}}          %  nu_2
\def\nuthree{\mbox{$\nu_3$}}        %  nu_3
\def\nus{\mbox{$\nu_s$}}            %  Sterile neutrino
\def\REarth{\mbox{$R_\oplus$}}   % Earth symbols
\def\MEarth{\mbox{$M_\oplus$}}
\def\BEarth{\mbox{$B_\oplus$}}
\def\RCore{\mbox{$R_{Core}$}}

\def\TYear{\mbox{$T_{y}$}}

\def\RhoR{\mbox{$\rho_R$}}      % Resonance density
\def\rhoR{\mbox{$\rho_R$}}      %     "        "

\def\Ye{\mbox{$Y_e$}}                 % Isotopic ratio
\def\YeCore{\mbox{$Y_e(Core)$}}       %    "       "   in the Core
\def\YeMantle{\mbox{$Y_e(Mantle$}}    %    "       "   in the Mantle

\def\ubar{\mbox{$\bar{u}$}}           % Atomic mass unit

\def\elmass{\mbox{$m_e$}}       % Electron mass

                                 % Energies / dms
\def\EDms{\mbox{$E_\nu/\Delta m^2$}} %   E/Dms
\def\Enu{\mbox{$E_\nu$}}         % Neutrino Energy
\def\Enucut{\mbox{$E_{\nu,cut}$}}% Neutrino cutting energy
\def\Te{\mbox{$T_e$}}            % Electron Kinetic energy
\def\TeTh{\mbox{$T_{e,th}$}}     % Threshold Electron energy
\def\Ee{\mbox{$E_e$}}            % Electron total energy
\def\EeTh{\mbox{$E_{e,th}$}}     % Threshold Electron total energy
                                 %

                                    % Neutrino Spectra
\def\SnuZr{\mbox{$\calS_0$}}        % Undistorted
\def\Snu{\mbox{$\calS$}}            % Distorted
                                    %

                                    % Electronic Spectra
\def\SeZr{\mbox{$\calS_{0}$}}       %    Undistorted 
\def\Se{\mbox{$\calS$}}             %    Distorted 
\def\Ses{\mbox{$\calS^s$}}          %       With s index
\def\SeD{\mbox{$\calS^D$}}          %       Day
\def\SeN{\mbox{$\calS^N$}}          %       Night
\def\SeC{\mbox{$\calS^C$}}          %       Core
\def\SeM{\mbox{$\calS^D$}}          %       Mantle
                                    %

                                    % Event Rate
\def\Re{\mbox{$\calR$}}             %   Generical
\def\Res{\mbox{$\calR^s$}}          %       "     with s index
\def\ReZr{\mbox{$\calR_{0}$}}       %   Solar Standard Model
\def\ReZrs{\mbox{$\calR_0^s$}}      %        "   "       "   with s index
\def\ReZrD{\mbox{$\calR_{0}^D$}}    %        "   "       "   for Day
\def\ReZrN{\mbox{$\calR_{0}^N$}}    %        "   "       "   for Night
\def\ReZrC{\mbox{$\calR_{0}^C$}}    %        "   "       "   for Core
\def\ReZrM{\mbox{$\calR_{0}^M$}}    %        "   "       "   for Mantle
                                    %
                                    %   Earth effect:
\def\ReD{\mbox{$\calR^D$}}          %     Day
\def\ReN{\mbox{$\calR^N$}}          %     Night
\def\ReC{\mbox{$\calR^C$}}          %     Core
\def\ReM{\mbox{$\calR^M$}}          %     Mantle
                                    %

                                    % Asymmetries
\def\Asym{\mbox{$\calA$}}           % Generical
\def\AsymP{\mbox{$\calA_P$}}        %   Probabilities
\def\AsymPs{\mbox{$\calA_P^s$}}     %      "     with s index
\def\AsymPN{\mbox{$\calA_P^N$}}     %      night
\def\AsymPC{\mbox{$\calA_P^C$}}     %      core
\def\AsymPM{\mbox{$\calA_P^M$}}     %      mantle
\def\AsymS{\mbox{$\calA_{D-N}$}}    %   Spectral
\def\AsymSs{\mbox{$\calA_{D-N}^s$}} %      "     with s index
\def\AsymSN{\mbox{$\calA_{D-N}^N$}} %      night
\def\AsymSC{\mbox{$\calA_{D-N}^C$}} %      core
\def\AsymSM{\mbox{$\calA_{D-N}^M$}} %      mantle
\def\AsymR{\mbox{$A_{D-N}$}}        %   Event Rates
\def\AsymRs{\mbox{$A_{D-N}^s$}}     %        "      with s index
\def\AsymRN{\mbox{$A_{D-N}^N$}}     %      night
\def\AsymRC{\mbox{$A_{D-N}^C$}}     %      core
\def\AsymRM{\mbox{$A_{D-N}^M$}}     %      mantle
\def\AsymRNCM{\mbox{$A_{D-N}^{N,C,M}$}}   %      night - core - mantle
\def\AsymRNM{\mbox{$A_{D-N}^{N(M)}$}}   %      night - mantle
                                    %

                                        % Spectral Distortion
\def\deltaS{\mbox{$\delta \calS$}}      %   Generical
\def\deltaSs{\mbox{$\delta \calS^s$}}   %   "      with s index
\def\deltaSD{\mbox{$\delta \calS^D$}}   %   Day
\def\deltaSN{\mbox{$\delta \calS^N$}}   %   Night
\def\deltaSC{\mbox{$\delta \calS^C$}}   %   Core
\def\deltaSM{\mbox{$\delta \calS^M$}}   %   Mantle
                                        %

                                        %
                                        % Probabilities
\def\Ps{\mbox{${\bar{P}}_\odot$}}       % Survival probability in the Sun
\def\APTot{\mbox{${\bar{P}}_\oplus$}}   % Averaged Ps
\def\PTot{\mbox{$P_{\oplus}$}}          % Total probability
\def\PTots{\mbox{$P^s_\oplus$}}         %     "             for sample s
\def\PeTw{\mbox{$P_{e2}$}}              % Transition nu_2 -> nu_e
\def\APeTw{\mbox{$<\PeTw>$}}            % Averaged Probability
\def\APeTws{\mbox{$<\PeTw>^s$}}         % Averaged Probability for sample s
\def\APeTwN{\mbox{$<\PeTw>^{N}$}}       %    Night
\def\APeTwC{\mbox{$<\PeTw>^{C}$}}       %    Core
\def\APeTwM{\mbox{$<\PeTw>^{M}$}}       %    Mantle
\def\APeTwDC{\mbox{$<\PeTw>^{DC}$}}     %    Deep Core

\def\TResid{\mbox{$T_{res}$}}         % Residence time
\def\TResids{\mbox{$T_{res}^s$}}      %   SAMPLE s
\def\TResidD{\mbox{$T_{res}^D$}}      %   Day
\def\TResidN{\mbox{$T_{res}^N$}}      %   Night
\def\TResidC{\mbox{$T_{res}^C$}}      %   Core
\def\TResidDC{\mbox{$T_{res}^{DC}$}}  %   Deep Core
\def\TResidM{\mbox{$T_{res}^M$}}      %   Mantle
                                      %

                                                 % Sampling nomenclature
\def\DAY{\mbox{\em{Day}}}                        %    Day
\def\night{\mbox{\em{Night}}}                    %    Night
\def\core{\mbox{\em{Core}}}                      %    Core
\def\mantle{\mbox{\em{Mantle}}}                  %    Mantle
\def\deepcore{\mbox{\em{Deep - Core}}}           %    Deep Core
\def\standard{\mbox{\em{Standard}}}              %    Standard
\def\hatdelta{\mbox{$\hat{\delta}$}}             % Sampling function
                                                 %

                                       % Solar coordinates
\def\alphaSun{\mbox{$\alpha_\odot$}}   %    Alpha
\def\deltaSun{\mbox{$\delta_\odot$}}   %    Delta
\def\RSE{\mbox{$R$}}                   % Solar - Earth distance
\def\Ellipt{\mbox{$\epsilon_0$}}       % Orbital ellipticity
                                       %
                                       % Elongations:
\def\hathmid{\mbox{$\hath_{m}$}}       %    midnight
\def\hathmmax{\mbox{$\hath_{m,max}$}}  %    minima
\def\hathmmin{\mbox{$\hath_{m,min}$}}  %    maxima
                                       %

                                       % Probabilities Peaks
\def\PeakM{\mbox{$M$}}                 %    Mantle
\def\PeakC{\mbox{$C$}}                 %    Core
\def\PeakCI{\mbox{$C_I$}}              %    Core I
\def\PeakCO{\mbox{$C_O$}}              %    Core O
                                       %
                                       % Asymmetries peaks
\def\PeakA{\mbox{$\calA$}}             %    A
\def\PeakHM{\mbox{$\calH^-$}}          %    H-
\def\PeakHP{\mbox{$\calH^+$}}          %    H+
                                       %

                                       % Cross Sections
%                                      % nue,e diff. cross section
\def\dseedEe{\mbox{$\frac{d\, \sigma_{\nu_e  } (\Te,E_\nu)}{d\,\Te}$}}
                                       %
%                                      % num,e diff. cross section
\def\dsemdEe{\mbox{$\frac{d\, \sigma_{\nu_\mu} (\Te,E_\nu)}{d\,\Te}$}}
                                       %

                                          % Other nomenclature
\def\daynight{D-N}                        % Day/Night expresion
\def\SK{Super - Kamiokande}
\def\maxim{\mbox{max}}                    % ``max''
\def\FORTRAN{\mbox{\tt{FORTRAN}}}
\def\MATHEMATICA{\mbox{\tt{MATHEMATICA}}}

\def\deg{\degres}  % degres symbol

                                 % Reaction Chains
\def\pp{\mbox{pp}}               %         "        pp
\def\pep{\mbox{pep}}             %         "        pep
\def\CNO{\mbox{CNO}}             %         "        CNO

\def\CdTv{\mbox{$\cos 2 \theta_V$}}
\def\CdTM{\mbox{$\cos 2 \theta_m$}}

\def\lambdaDtc{\mbox{$\lambda_D$}}

\def\NA{Nadir angle}

\draft
\begin{titlepage}
\preprint{\vbox{\baselineskip 10pt{
\hbox{Ref. SISSA 31/98/EP}
%\hbox{IASSNS -- AST 96/11}
\hbox{hep -- ph/9805262}
\hbox{8 May, 1998}}}}
\vskip -0.4cm
\title{ \bf Diffractive-Like (or Parametric-Resonance-Like?) Enhancement 
of the Earth (Day-Night) Effect \\
 for Solar Neutrinos Crossing the Earth Core}
\author{ S.T. Petcov $^{a,b)}$\footnote{Also at:
Institute of Nuclear Research and Nuclear Energy, Bulgarian Academy
of Sciences, BG--1784 Sofia, Bulgaria.}}
%\vglue -0.4cm
\address{a) Scuola Internazionale Superiore di Studi Avanzati, I-34013
Trieste, Italy}
\address{b) Istituto Nazionale di Fizica Nucleare, 
Sezione di Trieste, I-34013
Trieste, Italy}
\maketitle
\begin{abstract}
\begin{minipage}{5in}
\baselineskip 16pt

It is shown that the strong enhancement of the Earth (day-night) effect
for solar neutrinos crossing the Earth core in the case of the small mixing 
angle MSW $\nu_e \rightarrow \nu_{\mu(\tau)}$ transition solution 
of the solar neutrino problem is due to a new resonance 
effect in the solar neutrino transitions 
in the Earth and not just to the MSW effect in the core. 
The effect is in many respects similar to the electron
paramagnetic resonance.
The conditions for existence of this new resonance effect are discussed. 
They include specific constraints on the neutrino oscillation
lengths in the Earth mantle and in the Earth core, thus the
resonance is a ``neutrino oscillation length resonance''.
The effect exhibits strong dependence on the neutrino energy.
Analytic expression for the probability accounting for the solar 
neutrino transitions in the Earth, which provides a high precision 
description of the transitions, including the  
new resonance effect, is derived. The implications 
of our results for the searches of the day-night 
asymmetry in the solar neutrino experiments are also briefly discussed.
The new resonance effect is operative also in the 
$\nu_{\mu} \rightarrow \nu_{e}$ ($\nu_e \rightarrow \nu_{\mu}$)
transitions of atmospheric neutrinos crossing the Earth core.
 
\end{minipage}
\end{abstract}

\end{titlepage}

\newpage

\hsize 16.5truecm
\vsize 24.0truecm
\def\dm{$\Delta m^2$\hskip 0.1cm }
\def\dmsqua{$\Delta m^2$\hskip 0.1cm}
\def\sn{$\sin^2 2\theta$\hskip 0.1cm }
\def\snf{$\sin^2 2\theta$}
\def\trna{$\nu_e \rightarrow \nu_a$}
\def\trnm{$\nu_e \rightarrow \nu_{\mu}$}
\def\trns{$\nu_e \leftrightarrow \nu_s$}
\def\trnat{$\nu_e \leftrightarrow \nu_a$}
\def\trnmt{$\nu_e \leftrightarrow \nu_{\mu}$}
\def\trne{$\nu_e \rightarrow \nu_e$}
\def\trnst{$\nu_e \leftrightarrow \nu_s$}
\def\nue{$\nu_e$\hskip 0.1cm}
\def\numu{$\nu_{\mu}$\hskip 0.1cm}
\def\nutau{$\nu_{\tau}$\hskip 0.1cm}
\baselineskip 21.65pt
\font\eightrm=cmr8
\def\aprle{\buildrel < \over {_{\sim}}}
\def\aprge{\buildrel > \over {_{\sim}}}
\renewcommand{\thefootnote}{\arabic{footnote}}
\setcounter{footnote}{0}
\leftline{\bf 1. Introduction}
\vskip 0.2cm

\indent  In the present article we show that the strong 
enhancement of the Earth 
effect in the transitions of the solar 
neutrinos crossing the Earth core in comparison 
with the effect in the transitions of the 
only mantle crossing solar neutrinos, found \cite{Art1,Art2,Art3} to occur   
in the case of the matter-enhanced 
small mixing angle 
$\nu_e \rightarrow \nu_{\mu(\tau)}$
 solution of the solar neutrino problem,
is due to a new resonance 
effect in the transitions and not just to the MSW  enhancement 
of the neutrino mixing caused by the matter in the core.
The effect is in many respects analogous to the electron 
paramagnetic resonance.

   The existence of the core enhancement of the Earth matter effect
was established numerically in the rather detailed and high precision
studies of the day-night (D-N) effect for the Super-Kamiokande detector, performed
in \cite{Art1,Art2,Art3} for the 
$\nu_e \rightarrow \nu_{\mu(\tau)}$
and $\nu_e \rightarrow \nu_{s}$ transition
solutions. We have found, in particular, that due to this enhancement
the D-N asymmetry in the sample of events whose night fraction
is caused by the solar neutrinos which cross the Earth core before
reaching the detector, is in the case of the small mixing    
$\nu_e \rightarrow \nu_{\mu(\tau)}$ solution
by a factor of up to six bigger than the D-N asymmetry
determined using the whole night event sample produced by the solar neutrinos 
which cross only the Earth mantle or the mantle and the core. 
Such a strong enhancement was interpreted to be purely due
to the resonance in the neutrino transitions,
generated by the effect of the core matter
on the neutrino mixing and taking place
in the Earth core. We shaw in what follows that the origin of 
this enhancement is the presence of a 
new type of resonance
in the transitions of the solar neutrinos crossing the Earth core. 
For reasons to become clear later 
we will use the term ``neutrino oscillation length resonance'' for it. 

   In the analyzes which follow we use the Stacey 
model from 1977 \cite{Stacey:1977} as a 
reference Earth model. The Earth radius in the Stacey model 
is $R_{\oplus} = 6371~$km.
As in all Earth models known to us, the density distribution
is spherically symmetric and there are two major density structures - 
the core and the mantle, and 
a large number of substructures (shells or layers). The core 
has a radius $R_c = 3485.7~$km,
so the Earth mantle depth is approximately $R_{man} = 2885.3~$km.  
The mean matter densities in the core and in the mantle read, respectively:
$\bar{\rho}_c \cong 11.5~{\rm g/cm^3}$ and 
$\bar{\rho}_{man} \cong 4.5~{\rm g/cm^3}$.
Let us note that the density distribution in the 1977 Stacey model
practically coincides with the density distribution
in the more recent PREM model \cite{PREM81}.

 We will assume in the present study that the 
simplest two-neutrino 
$\nu_e - \nu_{\mu(\tau)}$ or $\nu_e - \nu_{s}$ 
mixing (with nonzero mass neutrinos),
$\nu_s$ being a sterile neutrino, 
takes place in vacuum and that it is at the origin of 
the solar $\nu_e$ matter-enhanced transitions into 
$\nu_{\mu(\tau)}$ or $\nu_{s}$ in the Sun, producing the observed
solar neutrino deficit. 

\vskip 0.2cm
\leftline{\bf 2. Enhancement of the Transitions
of Solar Neutrinos Crossing the Earth Core}
\vskip 0.2cm
 
 Because of the spherical symmetry of the Earth, the
path of a neutrino 
in the Earth mantle before the neutrino reaches the 
Earth core is identical to the path in the mantle after the neutrino 
exits the core; in particular, the lengths of the two 
paths in the mantle are equal. For the same reason a given 
neutrino trajectory is completely specified by its Nadir angle \cite{Art2}.

  It proves convenient to analyse the Earth  
effect in the transitions of 
solar neutrinos which cross the Earth core by using the two-layer model 
of the Earth: all the interesting
features of the transitions can be understood quantitatively 
in the framework of this rather simple model.
The density profile
of the Earth in the two-layer model 
is assumed to consist of two structures - 
the mantle and the core, 
having different densities, $\rho_{man}$ and $\rho_{c}$, and different
electron fraction numbers, $Y_e^{man}$ and $Y_e^{c}$, none of which however 
vary within a given structure. The core radius and the depth of the mantle
are known with a rather good precision and these data are incorporated
in the Earth models \cite{Stacey:1977,PREM81}.
The densities $\rho_{man}$ and $\rho_{c}$
in the case of interest should be considered as mean effective densities
along the neutrino trajectories, which can vary somewhat with
the change of the trajectory: $\rho_{man} = \bar{\rho}_{man}$ and 
$\rho_c = \bar{\rho}_{c}$.
In the Stacey model one has:
$\bar{\rho}_{man} \cong (4 - 5)~ {\rm g/cm^3}$ and 
$\bar{\rho}_{c} \cong (11 - 12)~ {\rm g/cm^3}$.
For the electron fraction numbers in the mantle and in the core one can use
the standard values \cite{Stacey:1977,PREM81,CORE} 
(see also \cite{Art1}) $Y_e^{man} = 0.49$ and $Y_e^{c} = 0.467$.
Numerical calculations show \cite{MP98:2layers} that, e.g., the 
time-averaged probability $P_{e2}$ calculated within 
the two-layer model of the Earth with $\bar{\rho}_{man}$
and $\bar{\rho}_{c}$ taken from
the Stacey 1977 model \cite{Stacey:1977} 
reproduces with a remarkably 
high precision the probability calculated 
by solving numerically the relevant system of evolution equations
with the much more sophisticated Earth density profile
of the Stacey model \cite{Stacey:1977}.

   The term which accounts for the Earth matter effect in 
the solar $\nu_e$ survival probability 
\footnote{This is the probability that
a solar $\nu_e$ will not be converted into $\nu_{\mu (\tau)}$ (or $\nu_s$)
when it travels from the central part of the Sun to the surface of the Earth
and further crosses the Earth to reach the detector.} 
in the case of interest is the probability of the 
$\nu_2 \rightarrow \nu_{e}$ transition in the Earth, $P_{e2}$,
where $\nu_2$ is the heavier of the two mass eigenstate
neutrinos in vacuum. 
The probability amplitude of 
interest $A(\nu_2 \rightarrow \nu_e)$,
$P_{e2} = |A(\nu_2 \rightarrow \nu_e)|^2$, is given by the following 
simple expression in the two-layer model:
$$A(\nu_2 \rightarrow \nu_e) = \sin\theta~ +~
\left( e^{-i2\Delta E'X'} - 1 \right) \left[ 1 + 
\left(e^{-i\Delta E''X''} - 1 \right) 
\cos^2(\theta'_{m} - \theta''_{m}) \right ]
\cos(\theta - \theta'_{m}) \sin\theta'_{m}~$$ 

\vspace*{-0.9cm}

$$ +~ \left( e^{-i\Delta E''X''} - 1 \right)
\cos(\theta - \theta''_{m})\sin\theta''_{m}~~~~~~~~~~~~~~~~~~~~~~$$

\vspace*{-1.0cm}

$$~~~~~~~~~~~~~~~~~~~~~~~~ +~ {1\over {2}} ~\left( e^{-i\Delta E''X''} - 1 \right)~
\left( e^{-i\Delta E'X'} - 1 \right) \sin(2\theta''_{m} - 2\theta'_{m})
\cos(\theta - 2\theta'_{m}). ~~~\eqno(1)$$

\noindent Here
$$\Delta E'~(\Delta E'') = 
\frac{\Delta m^2}{2E}
\sqrt{\left( 1 - \frac{\bar{\rho}_{man~(c)}}{\rho^{res}_{man~(c)}}
\right)^2 \cos^22\theta 
+ \sin^22\theta }~,~
\eqno(2)$$
\noindent $\theta'_{m}$ and  $\theta''_{m}$ are 
the mixing angles in matter in the
mantle and in the core, respectively,

$$\sin^22\theta'_{m}~(\sin^22\theta''_{m}) = \frac{\sin^22\theta}
{(1 - \frac{\bar{\rho}_{man(c)}}{\rho^{res}_{man(c)}})^2 \cos^22\theta + 
\sin^22\theta },~~~\eqno(3)$$

\noindent $X'$ is half of the distance the neutrino
travels in the mantle and $X''$ is the length of 
the path of the neutrino in the core,
and $\rho^{res}_{man}$ and $\rho^{res}_{c}$
are the resonance densities in the mantle and in the
core. The latter can be obtained from the expressions 
$$\rho^{res}_{a} = \frac{\Delta m^2 \cos2\theta}{2E\sqrt{2} G_F Y_e}~m_{N},~~~~~~~
\nu_e \rightarrow \nu_{\mu (\tau)}~~transitions,~~~\eqno(4)$$ 
$$\rho^{res}_{s} = 
\frac{\Delta m^2 \cos2\theta}{2E\sqrt{2} G_F \frac{1}{2}(3Y_e - 1)}~m_{N},~~~
\nu_e \rightarrow \nu_{s}~~transitions,~~~\eqno(5)$$ 

\noindent $m_{N}$ being the nucleon mass,
by using the
specific values of $Y_e$ in the mantle and in the core.
We have $\rho^{res}_{man}\neq \rho^{res}_{c}$  (but
$\rho^{res}_{man} \sim \rho^{res}_{c}$)
because $Y_e^{c} = 0.467$ and $Y_e^{man} = 0.49$.
Obviously, $Y_e^{c}\rho^{res}_{c} = Y_e^{man}\rho^{res}_{man}$.
For a neutrino trajectory which is specified by a given Nadir angle $h$,
the following relations hold true:
$$X' = R_{\oplus}\cosh -  \sqrt{R^{2}_{c} - R^2_{\oplus}\sin^2h},~~
X'' = 2\sqrt{R^{2}_{c} - R^2_{\oplus}\sin^2h}.~~\eqno(6)$$
 
  It is not difficult to find from eq. (1) that for any $\theta$ 
the probability of interest $P_{e2}$ has the form:
$$P_{e2} = \sin^2\theta~ +~ {1\over {2}} 
\left [1 - \cos \Delta E''X'' \right ] \left [ \sin^2(2\theta''_{m} - 
 \theta) -  \sin^2\theta \right ]~~~~~~~~~~~~~~~~~~~~~~~~~~~~~~~~~~~~~~~~~~~~~~~~$$

\vspace*{-1.0cm} 
$$ +~ {1\over {4}} \left [1 - \cos \Delta E''X'' \right ]
\left [1 - \cos \Delta E'X' \right ] 
\left [ \sin^2(2\theta''_{m} - 4\theta'_{m}  + \theta) - 
 \sin^2(2\theta''_{m} - \theta) \right ]~~~~~~~~~~ $$

\vspace*{-1.0cm}

$$ - ~{1\over {4}} \left [1 - \cos \Delta E''X'' \right ]
\left [1 - \cos 2\Delta E'X' \right ] 
\left [ \sin^2(2\theta'_{m} - \theta) - 
 \sin^2\theta \right ] \cos^2(2\theta''_{m} - 2\theta'_{m})~~~~~~ $$

\vspace*{-1.0cm}
 
$$ ~~+~{1\over {4}} \left [1 + \cos \Delta E''X'' \right ]
\left [1 - \cos 2\Delta E'X' \right ] 
\left [ \sin^2 (2\theta'_{m} - \theta) - 
 \sin^2\theta \right ]~~~~~~~~~~~~~~~~~~~~~~~~~~~~~ $$

\vspace*{-1.0cm}
 
$$ +~{1\over {2}} \sin \Delta E''X'' ~\sin 2\Delta E'X'  
\left [ \sin^2(2\theta'_{m} - \theta) - 
 \sin^2\theta \right ] \cos(2\theta''_{m} - 2\theta'_{m})~~~~~~~~~~~~~~~~~~~~ $$

\vspace*{-1.0cm} 
 
$$~~~~~ +~ {1\over {4}} \left [ \cos (\Delta E'X' - \Delta E''X'') 
- \cos (\Delta E'X' + \Delta E''X'') \right ]
 \sin (4\theta'_{m} - 2\theta)  
 \sin (2\theta''_{m} - 2\theta'_m).~ \eqno(7)$$  
 
  The effect of the core enhancement of the probability 
$P_{e2}$ was established to be dramatic \cite{Art1,Art2,Art3}
at small mixing angles, $\sin^22\theta \ltap 0.10$.
For $\sin^22\theta \ltap 0.10$, the probability 
$P_{e2}$, considered as a function
of $E/\Delta m^2$, or equivalently of the density 
parameter \cite{Art1,Art2,Art3}
$$\rho_r = \frac{\Delta m^2 \cos 2\theta}{2 E \sqrt{2} G_{F} C_{a(s)}}m_{N}~,~~\eqno(8)$$

\noindent where the constant $C_{a(s)} = 0.50~(0.25)$ for the
$\nu_e \rightarrow \nu_{\mu (\tau)}$ 
($\nu_e \rightarrow \nu_{s}$) transitions
\footnote{The parameter $\rho_r$ would coincide with the 
resonance density if \Ye\ were equal to
$1/2$ both in the mantle and in the core; it is equivalent to 
$E/\Delta m^2$, but gives an idea about the densities 
at which one has an enhancement of the
Earth matter effect and we have used 
it for the latter purpose in \cite{Art1,Art2,Art3}.}, 
can have several prominent local maxima in its bulk region.
When the MSW resonance 
(i.e., the resonance in $\sin^22\theta'_{m}$ or $\sin^22\theta''_{m}$)
takes place in the Earth core, 
we have $\rho^{res}_{c} \cong \bar{\rho}_{c}$, 
and correspondingly
$2\theta''_m \cong \pi/2$, while
$\theta'_m \cong  
1.7\theta \ll \theta''_m$. In the resonance region the probability
$P_{e2}$, as it follows from
eq. (7), is given by the expression:
$$P^{cres}_{e2} \cong   
{1\over {2}} 
\left [1 - \cos \Delta E''X'' \right ] \sin^2 (2\theta''_{m} - \theta)~,\eqno(9)$$

\noindent as long as the oscillating factor
$0.5(1 - \cos \Delta E''X'')$ is not small.
Obviously, $P^{cres}_{e2}$ can be suppressed even when
$\sin^22\theta'_{m} \cong 1$ if
in the resonance region of $\sin^22\theta''_{m}$ one has
$\cos \Delta E''X'' \cong 1$. In the latter case 
the other terms in eq. (7), notably the last one 
with the factor $\sin (4\theta'_{m} - 2\theta)~ 
\sin (2\theta''_{m} - 2\theta'_{m})$,
can give substantial contribution 
in $P_{e2}$ in the region where 
$\sin^22\theta''_{m}$ is strongly enhanced by the matter effect.
 
   If the MSW resonance occurs in the mantle, i.e., 
if $\rho^{res}_{man} \cong \bar{\rho}_{man}$, 
then $2\theta'_m \cong \pi/2$,  
$\pi/2 - \theta''_m < \theta \ll \theta'_m$ and 
at small mixing angles 
we expect $P_{e2}$ to be given
in the resonance region of $\sin^2 2\theta'_{m}$ 
approximately by the expression
$$P^{mres}_{e2} \cong {1\over {4}} \left [1 + \cos \Delta E''X'' \right ]
\left [1 - \cos 2\Delta E'X' \right ] 
\sin^2 2\theta'_{m}.\eqno(10)$$

\noindent The effect of the core will not suppress
the probability $P^{mres}_{e2}$ 
in this case only if $\cos \Delta E''X'' \cong 1$. 
However, if $\cos \Delta E''X'' \cong -1$
we get $P^{mres}_{e2} \cong 0$ in spite of the resonance 
in $\sin^2 2\theta'_{m}$ in the mantle.
The probability $P^{mres}_{e2}$ can also be suppressed if 
in the resonance region of $\sin^2 2\theta'_{m}$ one has, e.g.,
$(2\Delta E'X')^2 \ll 1$. In these cases $P_{e2}$
may turn out to be determined 
in the region of interest not by $P^{mres}_{e2}$, but rather 
by other terms in eq. (7) (see further).

  It is not difficult to convince oneself treating 
the phases $\Delta E'X'$ and $\Delta E''X''$ as independent 
parameters and $\theta$, $\theta'_m$ and
$\theta''_m$ as having fixed values
that for any $\theta$ 
the probability $P_{e2}$ has a local maximum   
if the following conditions are fulfilled:
\vspace*{-0.2cm}
$$\Delta E'X' = \pi (2k + 1),~~ \Delta E''X'' = \pi (2k' + 1), 
~~k,k' = 0,1,2,...~,~\eqno(11)$$
\noindent and 
$$\sin^2(2\theta''_{m} - 4\theta'_{m} + \theta) - \sin^2\theta > 0,
~~\eqno(12a)$$
$$\sin^2(2\theta''_{m} - 4\theta'_{m}  + \theta)\sin(2\theta''_{m} 
 - 2\theta'_{m})
\sin(2\theta'_{m} - \theta)\cos(2\theta''_{m} - 4\theta'_{m} + \theta)$$
$$ +~\frac{1}{4}\sin^2\theta \sin(4\theta''_{m} - 8\theta'_{m} + 2\theta)
\sin(4\theta''_{m} - 4\theta'_{m}) < 0.~~\eqno(12b)$$

\noindent Conditions (11) are written 
in the most general form. 
As we shall see, the relevant conditions for 
the problem of interest - transitions of solar 
neutrinos crossing the Earth core, correspond to $k = k' = 0$. 
At the maximum one has:
\vspace*{-0.2cm}
$$P^{max}_{e2} = \sin^2(2\theta''_{m} - 4\theta'_{m}  + \theta).~\eqno(13)$$

\noindent This local maximum will dominate in  
$P_{e2}$ at small mixing angles provided
$$P^{max}_{e2} >~(\gg)~max~P^{mres}_{e2},~max~P^{cres}_{e2}~,~~\eqno(14)$$

\noindent and if it is sufficiently wide. 
It is not clear a priori whether inequalities 
(14) will hold in the case of interest even if all the 
other conditions for the presence of the maximum in $P_{e2}$ are fulfilled.

   The requirements (12a) and (12b) are the two supplementary  
conditions which ensure that $P_{e2}$ has a maximum when
the equalities (11) hold
\footnote{If the sign of the inequality in eq. (12a) is opposite,
$P_{e2}$ would have a minimum.}. 
At small mixing angles condition (12a) is fulfilled for 
$\rho^{res}_{man} < 
(2\bar{\rho}_{c}Y_e^{c} /(\bar{\rho}_{man}Y_e^{man}) - 1)\bar{\rho}_{man} 
\cong  17.4~{\rm g/cm^3}$, $\rho^{res}_{man} \neq \bar{\rho}_{man}$,
and in this region condition (12b)
reduces to the following simple constraint
\footnote{It is not difficult to convince oneself that in 
the problem of interest one always has:
$\sin(2\theta''_{m} - 2\theta'_{m}) > 0$ and 
$\sin(2\theta'_{m} - \theta) > 0$.}:
$$\cos (2\theta''_{m} - 4\theta'_{m}  + \theta) < 0~~~\eqno(15)$$  

\noindent It can be shown that conditions
(12a) and (12b) (or condition (15)) can be satisfied if
$$\bar{\rho}_{man} < \rho^{res}_{man,c} < \bar{\rho}_{c},~~\eqno(16)$$

\noindent i.e., when $0 < 2\theta'_{m} < \pi/2$ 
and $\pi/2 < 2\theta''_{m} \leq \pi$.
They are always fulfilled, in particular, when  
$\bar{\rho}_{man} \ll \rho^{res}_{man}~( 2\theta'_{m} \cong 2\theta)$ and 
$\bar{\rho}_{c}/\rho^{res}_{c} \gg 1,\tan^22\theta~(2\theta''_{m} \cong \pi)$,
but these inequalities are not realized for the Earth as
$\bar{\rho}_{c} \cong 2.6\bar{\rho}_{man}$; moreover 
in this case $\sin^2(2\theta''_{m} - 4\theta'_{m}  + \theta) \cong
\sin^2(3\theta)$ and at small mixing angles $P^{max}_{e2}$ is suppressed. 
Actually, this observation indicates that the ratio between
$\bar{\rho}_{c}$ and $\bar{\rho}_{man}$ for the Earth is favorable
for the effect under discussion.
If, however, $\bar{\rho}_{man},\bar{\rho}_{c} < \rho^{res}_{man,c}$
($2\theta'_{m}, 2\theta''_{m} < \pi/2$, $2\theta'_{m} < 2\theta''_{m}$), 
the inequalities (12a) and (12b) will not hold. 
They do not hold also, e.g., for
$\bar{\rho}_{man},\bar{\rho}_{c} \ll \rho^{res}_{man,c}$, as well as when
$\rho^{res}_{man} = \bar{\rho}_{man}$, $\rho^{res}_{c} < \bar{\rho}_{c}$ or
$\rho^{res}_{man} > \bar{\rho}_{man}$, $\rho^{res}_{c} = \bar{\rho}_{c}$.
 
 If only $\Delta E'X' = \pi (2k + 1)$, one finds 
from eq. (7) that for any $\theta$
$$P_{e2} = \sin^2\theta~ +~ {1\over {2}} 
\left [1 - \cos \Delta E''X'' \right ] \left [ \sin^2(2\theta''_{m} - 
4\theta' + \theta) - \sin^2\theta \right ],~\Delta E'X' = \pi(2k + 1).~~\eqno(17)$$  
\noindent Note that at small mixing angles this expression 
differs from the one in eq. (9).
Similarly, when $\Delta E''X'' = \pi (2k' + 1)$, for any $\theta$ we obtain:
$$P_{e2} = \sin^2(2\theta''_{m} - \theta)
~+~{1\over {2}} \left [1 - \cos \Delta E'X' \right ] 
\left [ \sin^2(2\theta''_{m} - 4\theta'_{m}  + \theta) - 
 \sin^2(2\theta''_{m} - \theta) \right ]~~~~~~~~~$$

\vspace*{-1.0cm}

$$~- ~{1\over {2}} 
\left [1 - \cos 2\Delta E'X' \right ] 
\left [ \sin^2(2\theta'_{m} - \theta) - 
 \sin^2\theta \right ] \cos^2(2\theta''_{m} - 2\theta'_{m}),
~~\Delta E''X'' = \pi(2k' + 1).~\eqno(18)$$
 
  Let us discuss next the relevance and the implications of the above results
for the transitions of the solar neutrinos in the Earth.

\vglue 0.3cm
\leftline{\bf 2.1 MSW Resonance in the Earth Core}

  Consider first the case of $\nu_e \rightarrow \nu_{\mu (\tau)}$
transitions. If the MSW resonance takes place in the core
we have $\sin^22\theta_{m}''\cong 1$ for
$\rho^{res}_{c} \cong \bar{\rho}_{c} \cong (11.0 - 11.5)~ {\rm g/cm^3}$.  
However, the position of the 
maximum of $P_{e2} \cong P^{cres}_{e2}$ (eq. (9)) in the resonance region of
$\sin^22\theta_{m}''$ is determined by the condition 
$$\Delta E''X'' = \pi(2k' + 1),~k'=0,1,...,~~\eqno(19)$$

\noindent for which the oscillating term in eq. (9), 
$0.5(1 - \cos \Delta E''X'')$, has a maximal value.  
For $k = 0$ this condition  
is fulfilled for the core-crossing neutrinos for
$\rho^{res}_{c} \cong (13.3 - 14.5)~{\rm gr/cm^3}$.
Thus, the influence of the oscillating 
term in eq. (9) shifts the position of the 
maximum of $P^{cres}_{e2}$ 
to values of $\rho^{res}_{c}$ 
which differ from those at which 
the maximum of $\sin^22\theta_{m}''$ is located.
For the trajectory with
$h = 0^{0}$, for instance, the maximum of 
$P^{cres}_{e2}$ occurs at 
$\rho^{res}_{c} \cong 13.4~ {\rm g/cm^3}$ 
($(E/\Delta m^2)^{cres}_{a} \cong 1.053\times 10^{6}~{\rm MeV/eV^2}$), 
while for those with $h = 13^{0}$ and
$h = 23^{0}$ it takes place respectively at 
$\rho^{res}_{c} \cong 14.2~ {\rm g/cm^3}$
($(E/\Delta m^2)^{cres}_{a} \cong 0.993\times 10^{6}~{\rm MeV/eV^2}$)
and $\rho^{res}_{c} \cong 14.5~ {\rm g/cm^3}$
($(E/\Delta m^2)^{cres}_{a} \cong 0.975\times 10^{6}~{\rm MeV/eV^2}$). 
At small mixing angles the values of $\rho^{res}_{c}$ at which
$\Delta E''X'' \cong \pi$ exhibit weak dependence on 
$\sin^22\theta$. At the position of the maximum of 
$P^{cres}_{e2}$, $\sin^2 2\theta_{m}''$ is still considerably
enhanced by the matter effect:
for $\sin^22\theta = 0.01$, for instance,
on finds for $h = 13^{0};~23^{0}$ 
that at the maximum $\sin^2(2\theta_{m}'' - \theta) \cong % 0.26;~
0.15;~0.11$. 
For the trajectories with $h = 0^{0};~13^{0};~23^{0}$ we have
$max~P^{cres}_{e2} \cong 0.20;~0.15;~0.11$, in very good 
agreement with the results of our numerical calculations
\footnote{Let us note that for the indicated trajectories,
at the positions of the corresponding maxima of $P^{cres}_{e2}$ one has
$\Delta E'X' \cong 3\pi/2 ~(h = 0^{0})$, 
$\Delta E'X' \cong 1.6\pi ~(h = 13^{0})$ and
$\Delta E'X' \cong 2\pi ~(h = 23^{0})$.}  
\cite{Art1,Art2} (see Figs. 1a - 1c).

  If solar neutrinos undergo $\nu_e \rightarrow \nu_{s}$ transitions, 
the value of $E/\Delta m^2$ corresponding to a
given $\rho^{res}_{c}$ is for $Y_e^{c} = 0.467$, 
as it follows from eqs. (2) and (3),
by a factor of 2.33 larger than in the case of 
$\nu_e \rightarrow \nu_{\mu (\tau)}$ transitions: 
$(E/\Delta m^2)^{cres}_{s} \cong 2.33(E/\Delta m^2)^{cres}_{a}$.
Since $\Delta E'',\Delta E' \sim (E/\Delta m^2)^{-1}$, 
condition (20), $\Delta E''X'' \cong \pi$, is never satisfied
for the core-crossing neutrinos in, or in the vicinity of,  
the resonance region of $\sin^22\theta''_{m}$. Moreover, 
in the resonance region of interest   
$\Delta E''X''$ is relatively small
\footnote{At the point where $\sin^2(2\theta''_{m} - \theta) \cong 0.8$
($\rho^{res}_{c} \cong 10.9~ {\rm g/cm^3}$) for 
$\sin^22\theta = 0.01$, for instance, we have
$\Delta E''X''\cong 0.2\pi$.} 
and the oscillating term in eq. (9) suppresses the probability
$P^{cres}_{e2}$. As a consequence, $P_{e2}$ is given in the 
region of resonance enhancement of $\sin^22\theta''_{m}$ 
by the expression:
$$P^{cres}_{e2,s}\cong  
{1\over {2}} 
\left [1 - \cos \Delta E''X'' \right ] \sin^2 (2\theta''_{m} - \theta)
+~{1\over {4}} \left [1 + \cos \Delta E''X'' \right ]
\left [1 - \cos 2\Delta E'X' \right ] 
% \left [ 
\sin^2 2\theta'_{m}$$ %-  \sin^2\theta \right ]$$
% \vspace*{-1.0cm} 
$$+~{1\over {4}} \left [ \cos (\Delta E'X' - \Delta E''X'') 
- \cos (\Delta E'X' + \Delta E''X'') \right ]
 \sin (4\theta'_{m} - 2\theta)  
 \sin (2\theta''_{m} - 2\theta'_m).\eqno(20)$$ 

\noindent The maximal value of $P^{cres}_{e2,s}$ at
small mixing angles occurs for the 
trajectory with, e.g., $h = 0^{0}~(23^{0})$
at $\rho^{res}_{c} \cong 10.9~ {\rm g/cm^3}$
($\rho^{res}_{c} \cong 9.8~ {\rm g/cm^3}$)
and its position essentially does not 
change when $\sin^22\theta$ varies from 
$\sim 10^{-3}$ to $\sim 0.02$.
In the region of the maximum of $P^{cres}_{e2,s}$, 
$\Delta E'X'$ practically does not depend on
$\sin^22\theta \ltap 0.02$ and for 
$h = 0^{0}~(23^{0})$ we have at the maximum:
$\Delta E'X' \cong 0.40\pi ~(0.38\pi)$.
The value of the phase $\Delta E''X''$ in the region of the maximum 
of $P^{cres}_{e2,s}$ depends on
$\sin^22\theta$: if, e.g., $\sin^22\theta = 0.001;~0.01$ we get
for $h = 0^{0}~(23^{0})$ that at the maximum 
$\Delta E''X'' \cong 0.12\pi;~ 0.21\pi~(0.15\pi;~0.19\pi)$.
The mixing angle factors appearing in eq. (20)
have the following values at the maximum for, e.g.,  
$\sin^22\theta = 0.01$ and
$h = 23^{0}$: 
$\sin^2 (2\theta''_{m} - \theta) \cong 0.43$,
$\sin 4\theta'_{m} \cong 0.43$, and 
$\sin (2\theta''_{m} - 2\theta'_m) \cong 0.77$. 
The last term in eq. (20) is either the dominant 
one ($h \cong 23^{0})$ or is comparable in magnitude 
to the first term and their sum gives the dominant contribution in 
$P^{cres}_{e2,s}$ ($h \cong 0^{0})$. 
At the maximum the probability $P^{cres}_{e2,s}$
takes the following values for, e.g.,
$\sin^22\theta = 0.01$ and
$h = 0^{0};~23^{0}$: $max~P^{cres}_{e2,s} \cong 0.21;~0.15$,
which is in good agreement with the numerical results in ref. 
\cite{Art3} (see Figs. 2a - 2c).
It is interesting to note that 
for given $h$ and $\sin^22\theta$ one has: 
$max~P^{cres}_{e2,s} \sim max~P^{cres}_{e2}$.  

\vglue 0.3cm
\leftline{\bf 2.2 MSW Resonance in the Mantle}
 
 Suppose that for the core-crossing solar 
neutrinos the Earth matter effect leads to 
enhancement of $\sin^22\theta_{m}'$ due 
to the MSW resonance in the mantle. We shall assume first 
that the (solar) $\nu_e$ mix with $\nu_{\mu (\tau)}$ in vacuum.
In the region of the enhancement of $\sin^22\theta_{m}'$ 
the probability of interest $P_{e2} \cong P^{mres}_{e2}$ and 
is given approximately by eq. (10) as long as 
$P^{mres}_{e2}$ is not rather strongly suppressed.
Depending on the neutrino 
trajectory through the Earth core, we will have
$\sin^22\theta_{m}' \cong 1$ 
if $\rho^{res}_{m} \cong (4.0 - 5.0)~ {\rm g/cm^3}$.
However, the position of the maximum of $P^{mres}_{e2}$  
is determined by the position of the
maximum of the oscillating 
term $0.5(1 + \cos \Delta E''X'')$ in eq. (10),
i.e., by the condition
$$\Delta E''X'' = 2\pi(2k' + 1),~k'=0,1,...~,
~~\eqno(21)$$

\noindent and for most of the neutrino trajectories of interest
differs from the position of the maximum
of $\sin^22\theta_{m}'$. For 
$h = 0^{0}~(13^{0})$ and $h = 23^{0}$, for example, the
condition $\Delta E''X'' \cong 2\pi$ is realized 
at $\rho^{res}_{m} \cong 5.9~(5.1)~ {\rm g/cm^3}$
($(E/\Delta m^2)^{mres}_{a} \cong 2.270~(2.632)~\times 10^{6}~{\rm MeV/eV^2}$),
and at $\rho^{res}_{m} \cong 4.0~ {\rm g/cm^3}$
($(E/\Delta m^2)^{mres}_{a} \cong 3.374\times 10^{6}~{\rm MeV/eV^2}$). 

  Further, depending on the value of $\sin^22\theta$ and on the trajectory,
the maximum of $P^{mres}_{e2}$ can lie
outside or inside the resonance region of
$\sin^22\theta_{m}'$.
For the trajectories for which for given $\sin^22\theta \ltap 0.02$ 
the position of $max~P^{mres}_{e2}$ is within 
the resonance region of $\sin^22\theta_{m}'$ ($h\gtap 15^{0}$ if
$\sin^22\theta \cong 0.01$, for instance), we have $(\Delta E'X')^2 \ll 1$ 
at the maximum of $P^{mres}_{e2}$.
Correspondingly, the term $0.5(1 - \cos 2\Delta E'X')$
in the expression for $P^{mres}_{e2}$
suppresses $max~P^{mres}_{e2}$ even when both
$\Delta E''X'' \cong 2\pi$ and $\sin^2 2\theta'_{m} \cong 1$.
For the same reason 
$0.5(1 - \cos 2\Delta E'X')\sin^2 2\theta'_{m} \cong
(2\pi X'/L_v)^2 \sin^22\theta$ and practically 
does not depend on 
$\bar{\rho}_{m}$. When the maximum of  
$P^{mres}_{e2}$ lies outside the resonance region of
$\sin^22\theta_{m}'$, the term containing the factor
$\sin \Delta E''X'' ~\sin 2\Delta E'X'$ in eq. (7)
typically gives a non-negligible contribution
$\sim (15 - 30)\%$ to the value of 
$P^{mres}_{e2}$ at the maximum.
Even in the latter case $\sin^22\theta_{m}'$ is rather 
strongly enhanced at the maximum of
$P^{mres}_{e2}$.
For the trajectories corresponding to
$h = 0^{0};~13^{0};~23^{0}$ we get for 
$\sin^22\theta = 0.01$:
$max~P^{mres}_{e2} \cong 0.14;~0.10;~0.07$,
in very good agreement with our numerical results \cite{Art1,Art2}
(Figs. 1a - 1c). Note that at small mixing angles
$max~P^{mres}_{e2}$ is smaller than 
$max~P^{cres}_{e2}$ typically by a factor of $\sim 1.5$.

   If $\nu_e - \nu_{s}$ mixing takes place in vacuum,
a given value of $\rho^{res}_{m}$ corresponds now to a
2.09 larger value of $E/\Delta m^2$ than in the case of 
$\nu_e - \nu_{\mu (\tau)}$ mixing: 
$(E/\Delta m^2)^{mres}_{s} \cong 2.09(E/\Delta m^2)^{mres}_{a}$.
For this reason the condition 
$\Delta E''X'' \cong 2\pi$ is never satisfied
for the core-crossing neutrinos for values of
$\rho^{res}_{m}$ lying in, or relatively close, to the resonance region
of $\sin^22\theta_{m}'$. For $h = 0^{0}~(13^{0})$, 
$\sin^22\theta = 0.01$ and    
$\rho^{res}_{m} \cong 5.9~(5.1)~ {\rm g/cm^3}$, for instance,
$\Delta E''X'' \cong \pi$, and $P^{mres}_{e2}$ is suppressed.
For this value of $\rho^{res}_{m}$ we have
$(\Delta E'X')^2 \ll 1$ and $\Delta E'X'$ decreases with the
decreasing of $\rho^{res}_{m}$. As a consequence
of these circumstances, in the case of solar
$\nu_e \rightarrow \nu_{s}$ transitions,
the probability $P_{e2}$ at small mixing angles does not have a local maximum
in the region of the resonance enhancement of $\sin^22\theta'_{m}$:
as $\rho^{res}_{m}$ increases from 
$\sim 3.0~{\rm g/cm^3}$ to
$\sim 6.0~{\rm g/cm^3}$, $P_{e2}$
increases monotonically
\footnote{Lets us note that the Earth effect in the transitions
of the neutrinos crossing only the mantle 
can be understood very well qualitatively and quantitatively 
using the one-layer approximation 
for the density distribution in the Earth mantle
(see, e.g., ref. \cite{BaltzWen94}).}. 
This is well illustrated in Figs. 2a - 2c. 
 
\vglue 0.3cm
\leftline{\bf 2.3 Diffractive-Like Peak (or Parametric Resonance?) 
in $P_{e2}$: the Maximal Core}
\hskip 0.3truecm {\bf Enhancement of the Earth Effect}

 It is quite remarkable that for the 
$\nu_e \rightarrow \nu_{\mu (\tau)}$ solution 
of the solar neutrino problem
conditions (11) are approximately fulfilled 
and the maximal enhancement of  
the Earth effect in the
solar neutrino transitions at small mixing angles 
takes place for $\rho^{res}_{man,c}$ satisfying 
inequalities (16) and not in the regions of
the two possible MSW resonances (Figs. 1a - 1c).
Indeed, for, e.g.,  $\sin^22\theta = 0.01$ and the trajectories
corresponding to $h = 0^{0};~13^{0};~23^{0}$ 
we get $\Delta E'X' = \pi$ for 
$\rho^{res}_{m} \cong 10.0;~9.6;~8.0~{\rm g/cm^3}$
($(E/\Delta m^2)^{mres}_{a} 
\cong 1.646;~1.401;~1.317~\times 10^{6}~{\rm MeV/eV^2}$).
At the indicated values of $\rho^{res}_{m}$ the phase 
$\Delta E''X''$ is smaller than, but for most of the 
trajectories - rather close to, $\pi$. 
For the three trajectories indicated above we have:
$\Delta E''X'' \cong 0.55\pi;~ 0.75\pi;~0.88\pi$. 
When $\Delta E''X'' = \pi$, which takes place 
at smaller values of $\rho^{res}_{m}$, e.g.,  
in the example considered above
at $\rho^{res}_{m} \cong 8.7;~8.4;~7.6~{\rm g/cm^3}$,
the phase $\Delta E'X'$ is somewhat smaller than $\pi$
(in our example $\Delta E'X' \cong (0.75 - 0.90)\pi$). 
The same conclusions are valid for the other trajectories
of interest and values of $\sin^22\theta \ltap 0.02$.
Actually, the requirement that $\Delta E'X' = \Delta E''X'' = \pi$ 
is equivalent at small mixing angles in the 
case of interest 
to the condition
$$ \pi~\left [ {1\over {X'}} + {1\over {X''}} \right ] 
\cong \sqrt{2} {G_F\over{m_N}} (Y_e^{c}\bar{\rho}_{c} - 
Y_e^{man}\bar{\rho}_{man}).~~~\eqno(22)$$ 
  
\noindent It is not difficult to convince oneself using eq. (6), the values of
the core and Earth radii $R_{c}$ and $R_{\oplus}$ and the values of
$Y_e^{c}$, $\bar{\rho}_{c}$, $Y_e^{man}$ and $\bar{\rho}_{man}$ that the 
above equality cannot be exactly satisfied for the 
trajectories of neutrinos crossing the Earth core.

  The position of the absolute maximum of 
$P_{e2}$, which could be located, e.g., 
close to one of the two values of $\rho^{res}_{man}$ at which 
$\Delta E'X' = \pi$ or $\Delta E''X'' = \pi$,
is determined primarily by the properties of the 
function $\sin^2(2\theta''_{m} - 4\theta'_{m}  + \theta)$.
The latter has a minimum (although a relatively shallow one) 
at $\rho^{res}_{man} \sim 7.5 ~{\rm g/cm^3}$ 
at small mixing angles, 
and increases rather steeply monotonically 
when $\rho^{res}_{man}$ increases from 
$\sim 7.5~{\rm g/cm^3}$ to $\sim 11.5~{\rm g/cm^3}$. Since   
$\Delta E'X' = \pi$ occurs at larger values of $\rho^{res}_{man}$
than the equality $\Delta E''X'' = \pi$,
the position of the maximum of $P_{e2}$ of interest on the 
$\rho^{res}_{man}$ axis practically 
coincides with the position of the point where
$\Delta E'X' = \pi$. Correspondingly, the value of $P_{e2}$ at the maximum is
determined by the expression (17). For
$\sin^22\theta = 0.01$ and $h = 0^{0};~13^{0};~23^{0}$
we get  $max~P_{e2} \cong 0.51;~0.46;~0.40$, which is in 
beautiful agreement with 
the results of our numerical calculations \cite{Art1}
(see Figs. 1a - 1c). Note that, indeed, 
for a given trajectory through the core and given
$\sin^22\theta \leq 0.10$, 
$max~P_{e2}$ exceeds $max~P^{cres}_{e2}$ and $~max~P^{mres}_{e2}$ respectively 
by the factors of $\sim 2.5$ to $\sim 4.0$ and of $\sim 3.0$ to $\sim 7.0$. 
Moreover, because the MSW resonance in the core 
(and the corresponding maximum) is 
located relatively close
in $\rho^{res}_{man}$ (or $E/\Delta m^2$), the 
peak in $P_{e2}$ under discussion is rather wide
\footnote{Actually, as Fig. 1 indicates, for $\sin^22\theta \ltap 0.01$ and 
$0^{0} \leq h \ltap 15^{0}$ the MSW maximum appears just as a 
``shoulder'' on the slope of the maximum related to the 
conditions (11).}. In any case it is wider
than any of the local maxima corresponding to the 
MSW enhancement of the mixing at small mixing angles.  

  It should be noted also that it is the specific combination of 
mixing angles, $(2\theta''_{m} - 4\theta'_{m}  + \theta)$, and the fact that 
$\bar{\rho}_{c}$ and $\bar{\rho}_{man}$ do not differ by a large factor 
for the Earth, which makes $\sin^2(2\theta''_{m} - 4\theta'_{m}  + \theta)$ 
relatively large in the region of interest. The 
quantities $\sin^2 2\theta''_{m}$ and
$\sin^2 2\theta'_{m}$ associated with the MSW effect can be
actually rather small: if, for instance,
$\sin^22\theta = 0.01$, we find that at the position of the maximum of 
$P_{e2}$ under discussion, for the trajectory 
with $h = 23^{0}$ one has
$\sin^2 2\theta''_{m} \cong 0.08$ and
$\sin^2 2\theta'_{m} \cong  0.05$, while 
$\sin^2(2\theta''_{m} - 4\theta'_{m}  + \theta) \cong 0.40$.

   It is quite interesting that, as our analysis shows, for
values of $\rho^{res}_{man}$ from the interval
$(4.5 - 11.5)~{\rm g/cm^3}$, conditions (11) are not even approximately
fulfilled if solar neutrinos take part in $\nu_e \rightarrow \nu_{s}$ 
transitions, and hence 
the effect of strong enhancement of the probability
$P_{e2}$ considered above does not take place in this case.
However, as discussed in Section 2.1, when the  MSW 
resonance occurs in the core, the purely MSW-like term
(9) in $P_{e2}$, $P^{cres}_{e2}$,  
is ``assisted'' by some of the additional 
interference terms in $P_{e2}$ (see eq. (20)) and 
at the corresponding maximum, $P_{e2}$ is  by a factor of
$\sim (2 - 4)$ bigger than just $P^{cres}_{e2}$.
Still, at small mixing angles, $\sin^22\theta \ltap 0.02$, 
the value of $P_{e2}$ at its absolute maximum
in the $\nu_e \rightarrow \nu_{s}$ case is smaller than the
corresponding value when $\nu_e \rightarrow \nu_{\mu (\tau)}$
transitions take place by a factor of $\sim (2.5 - 4.0)$ 
(see also \cite{Art3}).

  As to the physical nature of this new type of enhancement of the probability
$P_{e2}$ which accounts for the Earth effect in the solar neutrino transitions,
it resembles the constructive interference of waves 
after diffraction, and the corresponding peak in  
$P_{e2}$ - a diffractive peak. 
This analogy (although not direct) is related to the 
fact that the amplitude of interest,
$A(\nu_2 \rightarrow \nu_e)$, has the following form in the two-layer model:
$$A(\nu_2 \rightarrow \nu_e) = \sum_{l,l'= e,\mu (\tau)} 
A^{man}(\nu_2 \rightarrow \nu_l)
A^{c}(\nu_l\rightarrow \nu_l')
A^{man}(\nu_l' \rightarrow \nu_e),~~~~\eqno(23)$$ 
 
\noindent where $A^{man}(\nu_2 \rightarrow \nu_l)$ and 
$A^{man}(\nu_l' \rightarrow \nu_e)$ are the probability
amplitudes of the $\nu_2 \rightarrow \nu_l$ and 
$\nu_l' \rightarrow \nu_e$ transitions in the Earth mantle 
(which is crossed twice by the neutrinos which traverse the Earth), while 
$A^{c}(\nu_l\rightarrow \nu_l')$ is the probability amplitude of 
$\nu_l\rightarrow \nu_l'$ transitions in the core, $l,l' = e,\mu (\tau)$.
It is the constructive interference of the different amplitudes
in the sum in eq. (23) which leads to the remarkable enhancement of
the probability $P_{e2}$ for the core-crossing neutrinos.

   Another possible interpretation of the enhancement of the probability
$P_{e2}$ discussed above is the existence of a parametric-like resonance
in the $\nu_2 \rightarrow \nu_e$ transitions in the Earth for which
the conditions eq. (11) are approximately satisfied, although the
term ``parametric-like'' does not suggest any specific analogy.
The possibility of a parametric-like
enhancement of the $\nu_e \rightarrow \nu_{\mu}$ transitions
in matter with density changing periodically but not continuously
along the neutrino path was considered in ref. \cite{Akh88}. 
Let us note that the case studied in
ref. \cite{Akh88} is different from the case considered by us.
Although in \cite{Akh88} 
one period in the variation
of density was assumed to consist of two layers 
with different finite densities (periodic step function), 
it was supposed that the two layers  
have equal spatial dimensions (widths) and that
the ``matter density is a periodic step function'', i.e. that
neutrinos cross an integer number of periods in density
while they propagate in matter. This does not correspond 
to the density profile of the Earth in the two-layer approximation.
Further, the results presented in \cite{Akh88}
were derived in the specific case 
of small vacuum mixing angle and for 
$\bar{\rho}_{man} \ll \rho^{res}_{man}$, 
$\bar{\rho}_{c} \ll \rho^{res}_{c}$. 
It should be clear from the discussion following eq. (16) 
that this specific case
also does not correspond to the one 
of transitions of neutrinos crossing the Earth core
studied by us here \footnote{In the earlier version of the present work
(hep-ph/9805262 of May 8, 1998)
we have incorrectly stated that conditions
(11) appear also in the specific case studied in 
ref. \cite{Akh88}.}. Finally, the relevant probability for 
the solar neutrino transitions in the Earth is the $\nu_2 \rightarrow \nu_{e}$ 
and not the $\nu_e \rightarrow \nu_{\mu}$ transition probability
which actually was considered in \cite{Akh88}. 

  There exists a beautiful analogy between the resonance effect
discussed in the present Section and the electron 
paramagnetic resonance taking place in a specific configuration
of magnetic fields \footnote{The author is grateful to L. Wolfenstein
who brought this analogy to the author's attention.}:
one constant, $\overrightarrow{B}_{0}$, located in the $xoz$ plane of the 
coordinate system and having a direction which forms an angle $2\theta$ 
with the z-axis,
$\cos2\theta > 0$, and a second, $\overrightarrow{B}_{1}$, 
along the z-axis, whose magnitude
can change step-wise in time. In the initial moment
$t = t_{0}$ the electron spin points up along the z-axis and 
in the interval of time $\Delta t_{1} = t_{1} - t_{0} > 0$ 
it is assumed to precess in the field $\overrightarrow{B}_{0}$ 
and in the constant field
$\overrightarrow{B}_{1}$ pointing down along 
the z-axis and having a relatively small magnitude
so that $B_{z} = B_{0}\cos2\theta - B_{1} > 0$, where $B_{0,1} = 
|\overrightarrow{B}_{0,1}|$. 
At $t = t_1$ the field $\overrightarrow{B}_{1}$ increases considerably (step-wise) 
in magnitude and in the interval of time $\Delta t_{2} = t_{2} - t_{1} > 0$
we have $B_{z} = B_{0}\cos2\theta - B_{1} < 0$. Finally, at
$t = t_{2}$, $\overrightarrow{B}_{1}$ changes (again step-wise) 
to its initial value 
and the precession of the electron spin continues
in such a configuration of fields for
\footnote{The last set of conditions is redundant
in the case of the electron paramagnetic resonance: we have added it
to make the analogy with the transitions of the 
neutrinos crossing the Earth core complete.} 
$\Delta t_{3} = t_{3} - t_{2} = \Delta t_{1}$. As can be shown,
there is one-to-one correspondence between the physical quantities
characterizing this problem and the ones in the problem
of transitions of neutrinos crossing the Earth core. 
At small values of $\sin^22\theta$ the probability
$P_{e2}$ corresponds to the probability
of the spin flip of the electron.
It is known that the latter can be made maximal ($\cong 1$)
by choosing $\Delta t_{1}$, $\Delta t_{2}$, $B_{0}$ and the two values of
$B_1$ in such a way that 
the precession angles in the intervals of time
$\Delta t_{1}$ and $\Delta t_{2}$ obey conditions
equivalent to (11) and the analog of the function
$\sin^2(2\theta''_{m} - 4\theta'_{m}  + \theta)$  for this problem
be close to 1. Obviously, in the case of neutrinos crossing the Earth core
the parameters are much more constrained because they are
predetermined by the properties of the Earth,
the limited interval of energies of solar neutrinos
and the rather small range of relevant values of $\Delta m^2$.
It seems almost a miracle
that conditions (11) can be fulfilled 
even approximately for the requisite
values of $\Delta m^2$ and $E$ and that the strong enhancement
of the probability $P_{e2}$, associated with 
the conditions (11), (12a) and (12b)
actually takes place.

  Since the existence of the resonance effect we have discussed
depends crucially on the conditions (11) which for a given neutrino 
trajectory through the Earth are actually conditions
on the neutrino oscillation length in the Earth core,
$L^{c} = 2\pi/\Delta E''$, and mantle,
$L^{man} = 2\pi/\Delta E'$, we shall use 
the term ``neutrino oscillation length resonance''
\footnote{In the earlier version of the present
article (hep-ph/9805262 of May 8, 1998) we have used conditionally the term
``diffractive-like enhancement'' for the new resonance effect. 
We think that ``neutrino oscillation length resonance''
better describes the essence of the effect.}. In contrast, 
the MSW effect is a resonance effect in the neutrino mixing.

   The implications of the oscillation length resonance enhancement of the
probability $P_{e2}$ for the core-crossing solar neutrinos
for the tests of the  
$\nu_e \rightarrow \nu_{\mu (\tau)}$ transition 
solution of the solar neutrino problem via the measurement
of the day-night effect related observables 
are discussed in detail in ref. \cite{Art1} (see also \cite{Art2,Art3}). 
It is quite remarkable that for values of
$\Delta m^2$ from the small (and part of the large)
mixing angle MSW solution region
the enhancement takes place for values of the 
$^{8}$B neutrino energy lying in the interval
$\sim (6 - 12)~$MeV to which the Super-Kamiokande and SNO 
experiments are sensitive. The peak in
$P_{e2}$ at $\sin^22\theta = 0.01$
for the trajectory with $h = 23^{0}$
taking place at $\rho^{res}_{man} \sim 8.0 ~{\rm g/cm^3}$, for instance,
corresponds to $E \cong 5.3~(10.5)~$MeV if 
$\Delta m^2 = 4.0~(8.0)\times 10^{-6}~{\rm eV^2}$. Correspondingly,
at small mixing angles this enhancement leads to 
a much bigger (by a factor of $\sim 6$) day-night asymmetry in the 
sample of events due to the core-crossing solar neutrinos
in the Super-Kamiokande detector
than the asymmetry determined by using the whole 
night event sample \cite{Art1}.
As a consequence, it may be possible to test a rather large part of the
small mixing angle solution region in the $\Delta m^2 - \sin^22\theta$
plane by performing selective day-night asymmetry measurements.
The new enhancement effect also leads 
to a rather large day-night asymmetry
in the recoil - e$^{-}$ spectrum which is being measured 
in the Super-Kamiokande experiment \cite{Art1}. 
It is not excluded that some of 
the future high statistics solar neutrino experiments will be able to 
observe directly this effect. A detector located at 
smaller geographical latitudes than the existing ones or 
those under construction \cite{Rosen97} would obviously be better suited
for this purpose.

\vglue 0.2cm
\leftline{\bf 3. Conclusions}
\vskip 0.2cm 
  We have shown in the present work that the strong 
enhancement of the Earth 
effect in the transitions of the solar 
neutrinos crossing the Earth core in comparison 
with the effect in the transitions of the 
only mantle crossing neutrinos 
in the case of the MSW 
small mixing angle 
$\nu_e \rightarrow \nu_{\mu(\tau)}$ transition 
solution of the solar neutrino problem \cite{Art1,Art2}, 
is due to a new oscillation length resonance 
effect in the transitions 
and not just to the MSW enhancement of the neutrino mixing
taking place in the Earth core.
The effect exhibits strong energy dependence. We have derived
analytic expression for the relevant transition
probability in the two-layer approximation for the
density distribution in the Earth, which reproduces
with high precision the probability calculated numerically by
using the Earth density profile
provided by the Earth model \cite{Stacey:1977}. The general 
conditions for the existence 
of the oscillation length resonance in 
the solar neutrino transitions in the
Earth were obtained and it was shown that at small 
mixing angles they are approximately
satisfied in the case of $\nu_e \rightarrow \nu_{\mu (\tau)}$ transitions.
These conditions are not fulfilled if solar neutrinos 
undergo $\nu_e \rightarrow \nu_{s}$ transitions.
Nevertheless, a similar but weaker (in what regards the 
maximal value of the relevant transition probability $P_{e2}$) 
enhancement takes place in the region of the MSW resonance 
in the core in the latter case.   
For solar neutrino transitions into active neutrino
and for the geographical
latitudes at which the Super-Kamiokande and SNO experiments
are located, the resonance enhancement takes place in the 
neutrino energy interval
$\sim (5 - 12)~{\rm MeV}$ if 
$\Delta m^2 \cong (4.0 - 8.0)\times 10^{-6}~{\rm eV^2}$, 
which is just in the region of the
small mixing angle MSW solution. If the transitions 
are into sterile neutrino it occurs at approximately 
two times larger neutrino energies for the same values of
$\Delta m^2$.

  The oscillation length resonance 
enhancement is present at large mixing angles 
in the transitions of solar neutrinos into 
an active neutrino \cite{Art1,Art2} as well.
It can also be present in the 
$\nu_{\mu} \rightarrow \nu_{e}$ (and $\nu_{e} \rightarrow \nu_{\mu}$) 
transitions of atmospheric neutrinos \cite{SPprep98}.
The probability of the $\nu_{\mu} \rightarrow \nu_{e}$ 
($\nu_{e} \rightarrow \nu_{\mu}$) transitions,
$P(\nu_{\mu} \rightarrow \nu_{e}) = P(\nu_{e} \rightarrow \nu_{\mu})$,
and the analogs of the maximum conditions (12a) and (12b) for this
probability (conditions (11) are the same as for $P_{e2}$)
can be obtained from eqs. (7), (12a) and (12b) by formally
setting $\theta = 0$ while keeping $\theta_{m}' \neq 0$ and 
$\theta_{m}'' \neq 0$. Thus, condition (12a) becomes
$\sin^2(2\theta''_{m} - 4\theta'_{m}) > 0$ and is always fulfilled,
while condition (12b) transforms into 
% $\cos(2\theta''_{m} - 4\theta'_{m}) < 0$.
$$\cos(2\theta''_{m} - 4\theta'_{m}) < 0.~~\eqno(24)$$
\noindent At small mixing angles, $\sin^22\theta \ltap 0.05$,
we have $P_{e2} \cong P(\nu_{\mu} \rightarrow \nu_{e})$
and thus Fig. 1 illustrates the dependence of
$P(\nu_{\mu} \rightarrow \nu_{e})$ on $\rho_r$ as well.
The new enhancement mechanism can be effective in this case
for $\rho^{res}_{c}$ and $\rho^{res}_{m}$ satisfying inequalities
(16). For $\sin^22\theta = 0.01$, 
$\Delta m^2 = 10^{-3}~(5\times 10^{-4})~{\rm eV^2}$, and center-crossing
($h = 0^{0}$), for instance, the absolute maximum of 
$P(\nu_{\mu} \rightarrow \nu_{e})$ takes place at 
$E \cong 1.6~(0.8)~$GeV.
Thus, for values of $\Delta m^2 \sim (5\times 10^{-4} - 5\times 10^{-3})~{\rm eV^2}$  
of the region of the $\nu_{\mu} \leftrightarrow \nu_{\tau}$ 
oscillation solution of the atmospheric neutrino problem the oscillation
length resonance strongly enhances the 
$\nu_{\mu} \rightarrow \nu_{e}$ (and $\nu_{e} \rightarrow \nu_{\mu}$)
transitions of the atmospheric neutrinos crossing the Earth core,
making the transition probabilities large 
(and perhaps the transitions detectable) 
even at small mixing angles.
The transitions under discussion should exist if 
three-flavour-neutrino mixing takes place in vacuum
\footnote{Actually, the excess of e-like events
in the region $-1 \leq \cos\theta_{z}\leq -0.6$,
$\theta_{z}$ being the Zenith angle, in the sub-GeV
sample of atmospheric neutrino events observed in the
Super-Kamiokande experiment (see, e.g., Y. Fukuda et al., hep-ex/9803006 
and submitted to Physics Letters B), can be due to 
$\nu_{\mu} \rightarrow \nu_{e}$ 
small mixing angle, $\sin^22\theta_{e\mu} \cong (0.01 - 0.10)$, transitions 
with $\Delta m^2 \sim 10^{-3}~{\rm eV^2}$,
strongly enhanced by the neutrino oscillation length
resonance as neutrinos cross the Earth core on the  way
to the detector. Such transitions are naturally predicted to exist
in a three-neutrino mixing scheme, in which, e.g., 
the small mixing angle MSW 
$\nu_{e} \rightarrow \nu_{\mu}$ transitions 
with $\Delta m^2_{21} \sim (4 - 8)\times 10^{-6}~{\rm eV^2}$,
or large mixing angle 
$\nu_{e} \leftrightarrow \nu_{\mu}$ oscillations
with $\Delta m^2_{21} \sim 10^{-10}~{\rm eV^2}$,
provide the solution of the solar neutrino
problem and the atmospheric neutrino anomaly is
due to $\nu_{\mu} \leftrightarrow \nu_{\tau}$
large mixing angle oscillations with 
$\Delta m^2_{31} \sim 10^{-3}~{\rm eV^2}$.
It should be added that if
$\Delta m^2_{31} \cong 
5\times 10^{-3}~{\rm eV^2}$, the 
excess of e-like 
atmospheric neutrino events in the Super-Kamiokande 
data due to  the neutrino oscillation length resonance should be present  
in the multi-GeV sample 
at $-1 \leq \cos\theta_{z}\ltap -0.8$.
}. 

 The new enhancement mechanism of $P(\nu_{\mu} \rightarrow \nu_{e})$
is operative at large mixing angles as well, at which the probabilities 
$P_{e2}$ and $P(\nu_{\mu} \rightarrow \nu_{e})$ exhibit different 
dependence on $E/\Delta m^2$ \cite{SPprep98}.

 The effects of the oscillation length resonance enhancement 
of the solar neutrino transitions in the Earth can be observed
in the currently operating high statistics solar neutrino experiments 
(Super-Kamiokande, SNO) by performing selective day-night asymmetry
measurements \cite{Art1}. A high statistics solar 
neutrino detector located at 
smaller geographical latitudes would be better suited for 
the observation of this enhancement. 

\newpage
%\vglue 0.4cm
\leftline{\bf Acknowledgements.}
The author would like to thank 
M. Maris for valuable discussions of the
day-night effect, V. Rittenberg for his kind and supportive interest in the
present work and B. Machet for helpful comments on the 
interpretation of the new effect suggested and discussed in this paper.
Thanks are also due to Zhenia Akhmedov, to whom the author
communicated the main results of the present work long before they were
assembled in the present publication, for graciously waiting for
the text of this article to be written in order to publish 
his own comments on the subject of the present study. 
This work was supported in part by the Italian MURST
under the program ``Fisica Teorica delle 
Interazioni Fondamentali''  and by Grant PH-510 from the
Bulgarian Science Foundation.

\vglue 0.5cm
% \newpage

\newpage

\bec{\Large\bf{Figure Captions}}\eec

\noindent
{\bf Figure 1.}
The dependence of the probability $P_{e2}$ on $\rho_r$ (see eq. (8)) 
in the case of $\nu_e \rightarrow \nu_{\mu (\tau)}$ transitions
of solar neutrinos for $\sin^22\theta = 0.01$ (from ref. \cite{Art2}). 
The five plots are obtained for 
$0.1 \mbox{ gr/cm}^3 \leq \rho_r \leq 30.0~\mbox{gr/cm}^3 $
($Y_e^{man} = 0.49$, $Y_e^{c} = 0.467$)
and five different solar neutrino trajectories in the Earth
determined by the Nadir angle $h$:
a) $h = 0^{0}$ (center crossing),
b) $h = 13^{0}$ (winter solstice for the Super-Kamiokande detector),
c) $h = 23^{0}$ (half core for the Super-Kamiokande detector),
d) $h = 33^{0}$ (core/mantle boundary),
e) $h = 51^{0}$ (half mantle).

\vspace{0.2cm}
\noindent
{\bf Figure 2.}
The dependence of the probability $P_{e2}$ on $\rho_r$ 
for $\nu_e \rightarrow \nu_{s}$ transitions
of solar neutrinos (from ref. \cite{Art3}). The five plots were obtained 
for $\sin^22\theta = 0.01$ and the same 
% range of values of $\rho_r$ and
neutrino trajectories as in figure 1.

\end{document}